\documentclass[aps, prd, amsmath, floats, floatfix, twocolumn,superscriptaddress, nofootinbib, showpacs,reprint]{revtex4-1}

\usepackage{graphicx}
\usepackage{amsmath,amssymb}
\usepackage{amsfonts}
\usepackage{xspace} 
\usepackage[usenames]{color}
\usepackage{dcolumn}
\usepackage{bm}
\usepackage{mathrsfs}

\usepackage{ulem}
\begin{document}

\title{Slowly rotating black holes in Ho\v rava--Lifshitz gravity}

\author{Enrico Barausse} 
\affiliation{Department of Physics, University of Guelph, Guelph, Ontario, N1G 2W1, Canada}
\affiliation{Institut d'Astrophysique de Paris, UMR 7095 du CNRS,
Universit\'{e} Pierre \& Marie Curie, 98bis Bvd. Arago, 75014 Paris, France}
\author{Thomas P. Sotiriou} 
\affiliation{SISSA, Via Bonomea 265, 34136, Trieste, 
Italy {\rm and} INFN, Sezione di Trieste}
\begin{abstract}
In a recent paper we claimed that there there are no slowly rotating, stationary, axisymmetric black holes in the infrared 
limit of Ho\v rava--Lifshitz gravity, provided that they are regular everywhere apart from the central singularity. Here we 
point out a subtlety in the relation between Einstein-$\ae$ther theory and the infrared 
limit of Ho\v rava--Lifshitz gravity which was missed in our earlier derivation and drastically modifies our conclusion: our 
earlier calculations (which are otherwise technically correct) do not really imply that there are no slowly rotating black holes 
in Ho\v rava--Lifshitz gravity, but that there are no slowly rotating black holes in the latter that are also solutions of Einstein-$\ae$ther 
theory and vice versa. That is, even though the two theories share the static, spherically symmetric solutions, there are no slowly rotating 
black holes that are solutions to both theories. We proceed to generate slowly rotating black hole solutions in the infrared limit of Ho\v rava--Lifshitz 
gravity, and we show that the configuration of the foliation-defining scalar remains the same as in spherical symmetry, 
thus these black holes are expected to possess a universal horizon.
\end{abstract}

\date{\today \hspace{0.2truecm}}

\maketitle

\section{Introduction}

In Ref.~\cite{ourpaper}  we considered  slowly rotating, stationary, axisymmetric black holes which are regular everywhere apart from the central singularity, and we claimed that such solutions do not exist in the infrared limit of Ho\v rava--Lifshitz (HL) gravity. This claim is actually incorrect. As we explain below, our derivation was implicitly using the relation between the infrared limit of HL gravity and Einstein-aether theory (\ae-theory). 
However, a subtle point in this relation has been missed, which effectively implies that we were implicitly requiring our solutions to be solutions of \ae-theory with a hypersurface-orthogonal aether configuration. The purpose of this note is threefold: a) to clarify how this oversight affected our calculations; b) to reinterpret our existing results, given that the calculations of Ref.~\cite{ourpaper} are actually technically correct and lead to some meaningful statements about \ae-theory; c) to correct our claims about HL gravity, point out that there are indeed slowly rotating black holes, and briefly discuss their characteristics.

In what follows we use the notation and definitions of Ref.~\cite{ourpaper} unless specified otherwise.

\section{A subtlety in the relation between HL gravity and $\ae$-theory}
Einstein-aether theory is described by the action
\begin{equation}
\label{actionae}
S_{\rm \ae}=\frac{M_{\rm \ae}}{2} \int d^4x \sqrt{-g}\left(-R-M^{\alpha\beta}_{\phantom{ab}\mu\nu} \nabla_\alpha u^\mu \nabla_\beta u^\nu \right),
\end{equation}
where $g$ is the determinant of the metric $g_{\mu\nu}$, 
$\nabla_\mu$ is the associated covariant derivative, 
$R$ is the Ricci scalar of this metric,
\begin{equation}
M^{\alpha\beta}_{\phantom{ab}\mu\nu}\equiv c_1 g^{\alpha\beta}g_{\mu\nu}+c_2 \delta^\alpha_\mu \delta^\beta_\nu+c_3 \delta^\alpha_\nu \delta^\beta_\mu+c_4 u^\alpha u^\beta g_{\mu\nu}\,,
\end{equation}
$c_1$ to $c_4$ are dimensionless parameters, and the aether $u^\mu$ is assumed to satisfy the constraint
\begin{equation}
u^\mu u_\mu=1\,.
\end{equation}
This constraint can be enforced either by considering restricted variations of the action which respect it, or by adding to the action explicitly the Lagrange multiplier $\zeta (u^\mu u_\mu -1)$.

It has been shown in  Ref.~\cite{Jacobson:2010mx} that $\ae$-theory is equivalent to the infrared limit of HL gravity if the aether is assumed to be hypersurface orthogonal {\em before} the variation. Locally, hypersurface orthogonality can be imposed through the condition
\begin{equation}
\label{hypers}
u_\mu=\frac{\partial_\mu T}{\sqrt{g^{\alpha\beta}\partial_\alpha T \partial_\beta T}}\,,
\end{equation}
where $T$ is a scalar field that defines a foliation. Choosing $T$ as the time coordinate one selects the preferred foliation of HL gravity, 
and the action (\ref{actionae}) reduces to the action of the infrared limit of HL gravity, whose Lagrangian is denoted as $L_2$ 
and given in eq.~(2) of Ref.~\cite{ourpaper}, and the 
correspondence of the parameters of the two theories is given in eq.~(6) of the same paper.

We consider now the corresponding field equations. Obviously, the equations of HL gravity will {\em not} be the equations of $\ae$-theory with the aether expressed as in eq.~(\ref{hypers}), as the equivalence only holds if the hypersurface-orthogonality condition is imposed at the level of the action. Let us consider the variation. Adding to action (\ref{actionae}) the Lagrange multiplier in order to enforce the unit constraint leads to
\begin{equation}
\label{aaef}
S=S_{\rm \ae}+\int d^4x \sqrt{-g} \,\zeta(u^\mu u_\mu-1)\,,
\end{equation}
and then variation yields
\begin{eqnarray}
\label{var1}
\delta S&=&\int d^4x\sqrt{-g} \Big[(E_{\mu\nu} +\zeta u_\mu u_\nu -\frac{1}{2} g_{\mu\nu} \zeta (u^\lambda u_\lambda -1))\delta g^{\mu\nu}\nonumber\\
&&\qquad \qquad+ (\AE^\mu+2\zeta u^\mu) \delta u_\mu+ (u^\lambda u_\lambda -1) \delta \zeta \Big],
\end{eqnarray}
where
\begin{equation}
E^{\mu\nu}\equiv \frac{\delta S_{\rm \ae}}{\delta g^{\mu\nu}},\qquad \AE^\mu\equiv  \frac{\delta S_{\rm \ae}}{\delta u_\mu}\,.
\end{equation}
Then the field equations are
\begin{eqnarray}
\label{ae1}
E_{\mu\nu} +\zeta u_\mu u_\nu -\frac{1}{2} g_{\mu\nu} \zeta (u^\lambda u_\lambda -1)&=&0\,,\\
\label{ae2}
\AE^\mu+2\zeta u^\mu&=&0\,,\\
\label{ucon}
u^\lambda u_\lambda &=&1\,,
\end{eqnarray}
being the $\ae$-theory field equation for the metric and the aether respectively. Contracting eq.~(\ref{ae2}) with $u_\mu$ and using the unit constraint, the equation can be re-written as
\begin{eqnarray}
\label{ae12}
E_{\mu\nu} -\frac{1}{2} \AE^\lambda u_\lambda u_\mu u_\nu&=&0\,,\\
\label{ae22}
\AE^\mu- \AE^\nu u_\nu u^\mu&=&0\,,\\
\label{ucon2}
u^\lambda u_\lambda &=&1\,.
\end{eqnarray}
These are the exact same equations one would obtain if, instead of the Lagrange multiplier, one had performed a restricted variation that implies the unit constraint (as in Ref.~\cite{Jacobson:2010mx}).

 If the hypersurface-orthogonality condition is imposed at the level of the action then the variation of $u_\mu$ has to be expressed in terms of $\delta T$ and $\delta g^{\mu\nu}$, i.e.
\begin{equation}
\delta u_\mu= -\frac{1}{2} u_\mu u_\nu u_\lambda \delta g^{\nu\lambda} +\frac{1}{\sqrt{g^{\alpha\beta}\partial_\alpha T \partial_\beta T}}(\delta^\nu_\mu -u^\nu u_\mu) \partial_\nu \delta T\,.
\end{equation}
 So, for (covariantized) HL gravity or hypersurface orthogonal $\ae$-theory one has
\begin{eqnarray}
\delta S_{h.o.}&=&\int \sqrt{-g} (E_{\mu\nu} \delta g^{\mu\nu} + \AE^\mu \delta u_\mu)\nonumber\\
&=&\int \sqrt{-g} \Big[ (E_{\mu\nu} -\frac{1}{2} \AE^\lambda u_\lambda u_\mu u_\nu) \delta g^{\mu\nu} \\
&&\qquad-\nabla_\nu\left( \frac{1}{\sqrt{g^{\alpha\beta}\partial_\alpha T \partial_\beta T}}(\delta^\nu_\mu -u^\nu u_\mu) \AE^\mu\right) \delta T\Big]\,,\nonumber 
\end{eqnarray}
which leads to the equations 
\begin{eqnarray}
\label{hl1}
E_{\mu\nu} -\frac{1}{2} \AE^\lambda u_\lambda u_\mu u_\nu&=&0\,,\\
\label{hl2}
\nabla_\nu\left( \frac{1}{\sqrt{g^{\alpha\beta}\partial_\alpha T \partial_\beta T}}(\delta^\nu_\mu -u^\nu u_\mu) \AE^\mu\right)&=&0\,.
\end{eqnarray}
The unit constraint is automatically satisfied, and this is why it is not present in the action or the field equations. 

It should be clear then that eq.~(\ref{hl1}) is identical to eq.~(\ref{ae12}) and hence, if one just starts with the ae-theory metric field equation and just imposes that the aether is given by eq.~(\ref{hypers}), then one obtains the HL gravity metric field equation. Obviously, the aether equation (\ref{ae22}) is different than the equation of motion for $T$, eq.~(\ref{hl2}). \footnote{Actually, it can be shown that diffeomorphism invariance (or the contracted Bianchi identity)
implies that eq.~(\ref{hl2}) follows directly from the Einstein equations (\ref{hl1})~\cite{Jacobson:2010mx}. }

Above we performed the variation keeping $u_\mu$ fixed in $\ae$-theory. Things become much more subtle if one attempts to keep $u^\mu$ fixed instead. In particular, starting from eq.~(\ref{var1}) and taking into account that
\begin{equation}
\delta u_\mu = g_{\mu\nu} \delta u^\nu - u_\nu g_{\mu \lambda} \delta g^{\nu\lambda}\,,
\end{equation}
one has
\begin{eqnarray}
\delta S_{\ae}&=&\int \sqrt{-g} \Big[(E_{\mu\nu} - \AE_{\mu}u_{\nu})\delta g^{\mu\nu}\nonumber\\
&&\qquad- \left(\zeta u_\mu u_\nu +\frac{1}{2} g_{\mu\nu} \zeta (u^\lambda u_\lambda -1)\right)\delta g^{\mu\nu}\nonumber\\
&&\qquad+ (\AE_\mu+2\zeta u_\mu) \delta u^\mu+ (u^\lambda u_\lambda -1) \delta \zeta \Big],
\end{eqnarray}
where $E_{\mu\nu}$ and $\AE^\mu$ are the same quantities as above. Then the equations are
\begin{eqnarray}
\label{ae1u}
E_{\mu\nu} - \AE_{(\mu}u_{\nu)}-\zeta u_\mu u_\nu -\frac{1}{2} g_{\mu\nu} \zeta (u^\lambda u_\lambda -1)&=&0\,,\\
\label{ae2u}
\AE_\mu+2\zeta u_\mu&=&0\,,\\
\label{uconu}
u^\lambda u_\lambda &=&1\,,
\end{eqnarray}
and eliminating $\zeta$ leads to
\begin{eqnarray}
\label{ae1u2}
E_{\mu\nu} - \AE_{(\mu}u_{\nu)}+\frac{1}{2} \AE^\lambda u_\lambda u_\mu u_\nu&=&0\,,\\
\label{ae2u2}
\AE^\mu- \AE^\nu u_\nu u^\mu&=&0\,,\\
\label{uconu2}
u^\lambda u_\lambda &=&1\,.
\end{eqnarray}
It is straightforward to see that eq.~(\ref{ae1u2}) differs from eq.~(\ref{ae12}). Of course, this does not mean that the dynamics of $\ae$-theory depends on what is kept fixed in the variation. One can use eq.~(\ref{ae2u2}) to show that
\begin{equation}
\AE_{(\mu}u_{\nu)}=\AE^\lambda u_\lambda u_\mu u_\nu \,.
\end{equation}

However, this subtlety is crucial for the relation with HL gravity. In particular, one cannot start from eq.~(\ref{ae1u2}), assume that the aether is given by eq.~(\ref{hypers}) and obtain eq.~(\ref{hl1}), 
as was possible when the $\ae$-theory equations were obtained with $u_\mu$ kept fixed. This was exactly what was missed in Ref.~\cite{ourpaper}, where we used in our calculation eq.~(\ref{ae1u2}) and (\ref{hypers}) instead of eqs.~(\ref{ae12}) and (\ref{hypers}), as an equivalent to eq.~(\ref{hl1}).

\section{Re-interpretation of earlier results}

When the aether is hypersurface orthogonal, eq.~(\ref{hl1}) and eq.~(\ref{ae1u2}) differ only by terms that vanish when the aether's equation, eq.~(\ref{ae2u2}) is 
satisfied. This implies that by considering in Ref.~\cite{ourpaper} the system of eqs.~(\ref{ae1u2}), (\ref{hypers}) and (\ref{hl2}), 
we effectively considered solutions of $\ae$-theory for which the aether is hypersurface orthogonal (at the level of the solution). Therefore, our 
earlier results imply that no such solutions exist, and $\AE$-theory can only admit slowly rotating black hole solutions where the aether is {\em not} hypersurface 
orthogonal. In HL gravity, instead, $u_\mu$ is by construction a hypersurface orthogonal vector. This implies that, even though the two theories share the static, 
spherically symmetric black hole solutions \cite{arXiv:1007.3503,ourpaper}, they do not share any slowly rotating solution.

Another interesting implication of the fact that $\ae$-theory does not admit slowly rotating solutions where the aether is hypersurface orthogonal is that slowly 
rotating black hole spacetimes in $\ae$-theory do not have a preferred foliation (but just a preferred frame locally). The existence of the universal horizons found in 
Refs.~\cite{Barausse:2011pu,Blas:2011ni} for static, spherically symmetric black holes appears to be related with the existence of a preferred foliation.

\section{Slowly rotating black holes in HL gravity}

Having re-interpreted the results of Ref.~\cite{ourpaper}, it should now be clear that the claim that there are no slowly rotating black holes in HL gravity is not supported. So, 
in the rest of the paper we re-address the question of finding slowly rotating solutions in HL gravity.

As discussed in Ref.~\cite{ourpaper} the most general ans\"atze for the metric and $u_\mu$ for a slowly rotating, stationary, axisymmetric black hole are without loss of generality\footnote{Here $u_\mu$ is given in terms of $T$ by eq.~(\ref{hypers}) but it is still more convenient to discuss the ansatz in terms of $u_\mu$.}
\begin{eqnarray}\label{metric}
ds^2&=&f(r) dt^2 -\frac{B(r)^2}{f(r)}dr^2-r^2(d\theta^2+\sin^2\theta \,d\varphi^2)\nonumber\\
&&+\epsilon 
r^2 \sin^2\theta \,\Omega(r,\theta) dtd\varphi+{\cal O}(\epsilon^2)\,,\\
\label{aether}
\boldsymbol{u}&=& \frac{1+f A^2}{2 A}dt
+ \frac{B}{2A}\left(\frac{1}{f}-A^2\right) dr+{\cal O}(\epsilon)^2\
\end{eqnarray}
where $\epsilon$ is the book-keeping parameter  of the expansion in the rotation, and $A(r)$, $B(r)$ and $f(r)$ are given by the static, spherically symmetric seed solutions
(see. Ref.~\cite{Barausse:2011pu}). Note in particular that hypersurface orthogonality implies $u_\varphi=0$ (at least to ${\cal O}(\epsilon)$) \cite{ourpaper}.

In  Ref.~\cite{ourpaper} we found it convenient  to perform a field
redefinition to the spin-0 metric 
$g'_{\alpha\beta}=g_{\alpha\beta}+(s_0^2-1)u_\alpha u_\beta$
 ($s_0$ being the speed of the spin-0 mode) and to the rescaled aether vector
$u'^\alpha=s_0^{-1} u^\alpha$~\cite{Eling:2006ec} and perform the analysis in terms of the redefined fields. 
The analysis for HL gravity actually turns out to be slightly simpler in the original 
variables $g_{\alpha\beta}$ and $u^{\mu}$, so we will avoid the field redefinitions here.
The key difference from the analysis of  Ref.~\cite{ourpaper}  is
 the issue discussed above. As a result,
the $r\phi$ and $t\phi$ components of the Einstein equations
get modified and become
\begin{align}\label{eq1}
&k_0\Omega+\frac{{c}_{13} \left(A^4 f^2-1\right)}{8
   r^2 A^2 B f}(\partial^2_\theta\Omega+3 \cot\theta \partial_\theta\Omega) =0\\
& \left(\partial^2_\theta\Omega+3 \cot\theta \partial_\theta\Omega\right) \frac{\left({c}_{13} A^4 f^2+2 A^2
   ({c}_{13}-2) f+{c}_{13}\right)}{8 A^2 f}\nonumber\\
   &+q_0\Omega-\frac{r ({c}_{13}-1) f \left(r B'-4 B\right)
   \partial_r\Omega}{2 B^3}\nonumber\\&+\frac{r^2 ({c}_{13}-1) f \partial_r^2\Omega}{2 B^2}=0\label{eq2}
\end{align}
which replace eqs. (12) and (13) of our paper.
Equation (11) of  Ref.~\cite{ourpaper}, which is the $\theta\varphi$ component of
the Einsten equations, instead remains unchanged at linear order in $\epsilon$ and is still given
by
 \begin{align}
 \label{eq0}
&\frac{{c}_{13}}{8
    r^3 A^3 B f^2} 
 \Big\{f \Big[2 \partial_\theta\Omega (r,\theta) (A- r A')
+r A   \partial_r\partial_\theta\Omega (r,\theta)\Big]\nonumber\\&
 \quad -f^3 A^4 \Big[2  \left(r A'+A\right) 
 \partial_\theta\Omega (r,\theta)+r A \partial_r\partial_\theta\Omega (r,\theta)\Big]\nonumber\\&
  \quad -r A f' \partial_\theta\Omega (r,\theta) \Big(1+A^4 f^2\Big)\Big\}=0\,.
\end{align}
(Also, note that the equation of motion of $T$, eq.~\eqref{hl2}, is identically satisfied at
linear order in $\epsilon$.)

 Just as in  Ref.~\cite{ourpaper}, here $k_0$ and $q_0$
are complicated functions of the couplings ${c}_i$,
 as well as of $A$, $f$ and $B$ and their derivatives,
but they must evaluate identically to zero
when one uses
the spherically symmetric static solution,
because $\Omega=$const must be a solution to the field equations~\cite{ourpaper}.
Combining eqs.~\eqref{eq1}--\eqref{eq2} we then immediately obtain
\begin{equation}\label{combo}
-\frac{r \left(r B'-4 B\right)
   \partial_r\Omega}{2 B^3}+\frac{r \partial_r^2\Omega}{2 B^2}=0\,,
\end{equation}
while from eq.~\eqref{eq0}, as in Ref.~\cite{ourpaper}, one can conclude that
if the black-hole horizon is to be regular and located at $r=r_{\rm H}$, 
one must have $\Omega(r_{\rm H},\theta)$= constant.
Alternatively, one can observe that  $\Omega(r,\theta)= \Omega(r)$ is the only solution to eq.~\eqref{eq1} that is regular at the poles
$\theta=0, \pi$.

We can then integrate eq.~\eqref{combo} and obtain
\begin{equation}\label{rot_soln}
\Omega(r,\theta)=\Omega(r)=- 12 J\int_{r_{\rm H}}^r \frac{B(\rho)}{\rho^4} d\rho +\Omega_0
\end{equation}
where $J$ and $\Omega_0$ are integration constants. In particular, because with asymptotically flat boundary conditions one has $B\sim 1$
far from the black hole, a comparison to the slowly
rotating Kerr metric highlights that $J$ plays the role of the spin of the black hole, while $\Omega_0$
can be eliminated from the metric with a coordinate change $\phi'=\phi-\Omega_0 t/2$.

\section{Conclusions}
Our claim, in Ref.~\cite{ourpaper}, that no slowly rotating regular black-hole solutions exist in HL gravity was incorrect. Such 
solutions actually exist and are given by eq.~\eqref{rot_soln}, as our amended analysis indicates. It goes beyond the scope of this manuscript to study the characteristics of these slowly rotating black holes, their deviations from the slowly rotating Kerr spacetime and any related astrophysical implications. We will address this question in a separate publication. It is, however, obvious, that the existence of these solutions implies that there is no
\textit{a priori} tension between the prediction of HL gravity and astrophysical evidence for the existence of spinning black holes, in contrast with our previous claim.

It is worth noting that the configuration of the foliations defining scalar $T$ (or $u_\mu$) in these solutions receives no correction at first order in the rotation, and hence it is effectively the same as in their spherically symmetric seed solution. Hence, we expect these solutions to possess a universal horizon.

The results found in Ref.~\cite{ourpaper}, when appropriately reinterpreted, imply that the slowly rotating solutions of HL gravity are not solutions of \ae-theory and that the latter has no slowly rotating solution with a hypersurface-orthogonal aether configuration. Hence, \ae-theory does not have slowly rotating black holes with a preferred foliation. Even though the two theories share the static, spherically symmetric, asymptotically flat solutions, they do not share slowly rotating solutions.

\acknowledgments{We are indebted to Ted Jacobson for private communications which contributed critically to understanding the subtlety
in the relation between the two theories and its implications, discussed in the first section of the manuscript. 
EB acknowledges support from a CITA National Fellowship
while at the University of Guelph, and from
the European Union's Seventh Framework Programme (FP7/PEOPLE-2011-CIG) through the
 Marie Curie Career Integration
Grant GALFORMBHS PCIG11-GA-2012-321608 while at the Institut d'Astrophysique de Paris. T.P.S. acknowledges financial support from the European Research Council under the European Union's Seventh Framework Programme (FP7/2007-2013) / ERC Grant Agreement n° 306425 ``Challenging General Relativity", from 
the Marie Curie Career Integration Grant LIMITSOFGR-2011-TPS Grant Agreement n° 303537, and from the ``Young SISSA Scientist Research project'' scheme 2011-2012.}

\end{document}